\documentclass[journal,twoside]{IEEEtran}

\usepackage{cite}
\usepackage{amsmath,amssymb,amsfonts}
\usepackage{graphicx}
\usepackage{hyperref}
\begin{document}
\title{Spectral Coherence Index: A Model-Free Metric for Protein Structural Ensemble Quality Assessment}
\author{Yuda~Bi, Huaiwen~Zhang, Jingnan~Sun, and Vince~D.~Calhoun,~\IEEEmembership{Fellow,~IEEE}
\thanks{Manuscript received XXXX XX, 2026.}
\thanks{Y.~Bi is with the Tri-institutional Center for Translational Research in Neuroimaging and Data Science (TReNDS), Georgia State University, Georgia Institute of Technology, and Emory University, Atlanta, GA 30303 USA (e-mail: ybi3@gsu.edu).}
\thanks{H.~Zhang is with the School of Computer Science and Technology, Xidian University, Xi'an 710126, China.}
\thanks{J.~Sun is with the Departments of Biomedical Engineering, Johns Hopkins University, Baltimore, MD 21287 USA.}
\thanks{V.~D.~Calhoun is with the Tri-institutional Center for Translational Research in Neuroimaging and Data Science (TReNDS), Georgia State University, Georgia Institute of Technology, and Emory University, Atlanta, GA 30303 USA, and also with the School of Electrical and Computer Engineering, Georgia Institute of Technology, Atlanta, GA 30332 USA (e-mail: vcalhoun@gsu.edu).}}

\maketitle

\begin{abstract}
Protein structural ensembles from NMR spectroscopy capture biologically important conformational heterogeneity, but it remains difficult to determine whether observed variation reflects coordinated motion or noise-like artifacts. We evaluate the Spectral Coherence Index (SCI), a model-free, rotation-invariant summary derived from the participation-ratio effective rank of the inter-model pairwise distance-variance matrix. Under grouped primary analysis of a Main110 cohort of 110 NMR ensembles (30--403 residues; 10--30 models per entry), SCI separated experimental ensembles from matched synthetic incoherent controls with AUC-ROC $= 0.973$ and Cliff's $\delta = -0.945$. Relative to an internal 27-protein pilot, discrimination softened modestly, showing that pilot-era thresholds do not transfer perfectly to a larger, more heterogeneous cohort: the primary operating point $\tau = 0.811$ yielded 95.5\% sensitivity and 89.1\% specificity. PDB-level sensitivity remained nearly unchanged (AUC $= 0.972$), and an independent 11-protein holdout reached AUC $= 0.983$. Across 5-fold grouped stratified cross-validation and leave-one-function-class-out testing, SCI remained strong (AUC $= 0.968$ and $0.971$), although $\sigma_{R_g}$ was the stronger single-feature discriminator and a QC-augmented multifeature model generalized best (AUC $= 0.989$ and $0.990$). Residue-level validation linked SCI-derived contributions to experimental RMSF across 110 proteins and showed broad concordance with GNM-based flexibility patterns. Rescue analyses showed that Main110 softening arose mainly from size and ensemble normalization rather than from loss of spectral signal. Together, these results establish SCI as an interpretable, bounded coherence summary that is most useful when embedded in a multimetric QC workflow for heterogeneous protein ensembles.
\end{abstract}


\section{Introduction}
\label{sec:introduction}

\IEEEPARstart{P}{roteins} are not static entities. Their biological
function---including catalysis, signal transduction, and molecular
recognition---depends critically on conformational dynamics spanning
timescales from picoseconds to seconds~\cite{henzler2007dynamic,
frauenfelder1991energy}. Nuclear magnetic resonance (NMR) spectroscopy
remains a primary experimental technique for characterizing this
heterogeneity at atomic resolution, producing multi-model structural
ensembles that represent conformations accessible to a protein in
solution~\cite{bonvin1995nmr, clore2004accurate, montelione2013recommendations}.
In parallel, modern deep-learning predictors achieve remarkable accuracy
for single-structure modeling, and recent efforts extend these models
toward sampling multiple conformations~\cite{jumper2021highly, baek2021accurate, lin2023evolutionary,
wayment2024predicting, del2022sampling, jing2024alphafold}.
Despite this progress, a fundamental gap persists: there is no widely
adopted, model-free metric that quantifies \emph{how coherently} an
ensemble’s structural variation is organized---i.e., whether inter-residue
distance changes are concentrated into a few coordinated modes or instead
appear as diffuse, noise-like fluctuations.

Several metrics characterize ensemble diversity, each with notable
limitations. Pairwise RMSD requires structural superposition and yields
a distribution rather than a single ensemble-level summary~\cite{diamond1992relationship, kufareva2012rmsd}.
Global measures such as the radius-of-gyration standard deviation
($\sigma_{R_g}$) miss local rearrangements or hinge-like motions that
preserve overall size~\cite{lobanov2008radius}. PCA-based variance ratios
depend on coordinate alignment and can be sensitive to the chosen
reference~\cite{amadei1993essential, david2014principal}, while contact-
or clustering-based summaries introduce arbitrary cutoffs or thresholds
that can yield discontinuous behavior~\cite{vendruscolo2002three, li2006cd}.
These limitations motivate a metric that (i)~operates directly on
inter-residue distances to avoid superposition artifacts; (ii)~summarizes
the spectral organization of distance variation rather than relying on a
single dominant component; and (iii)~produces a single interpretable
value on $[0,1]$.

Spectral methods have a long history in structural biology. Elastic
network models such as the Gaussian Network Model
(GNM)~\cite{bahar1997direct, haliloglu1997gaussian} and the Anisotropic
Network Model (ANM)~\cite{atilgan2001anisotropy} use the eigenspectrum
of a connectivity Laplacian or Hessian to predict collective motions
from a single structure. The concept of effective rank---quantifying the
effective dimensionality of a spectrum via the participation
ratio~\cite{vershynin2018high} or spectral
entropy~\cite{roy2007effective, scheurer2015exploration}---is widely used in signal
processing and random matrix theory to summarize spectral concentration
versus dispersion. Scheurer et al.~\cite{scheurer2015exploration} applied
spectral entropy to coordinate-covariance eigenspectra to assess
\emph{sampling completeness} in molecular dynamics trajectories.
In molecular dynamics more broadly, quasi-harmonic analysis and
principal component analysis of trajectories~\cite{garcia1992large,
amadei1993essential, balsera1996principal} decompose conformational
fluctuations into modes, but they operate on superposed Cartesian
coordinates and therefore depend on alignment and reference choices. To
our knowledge, no effective-rank measure has previously been applied to
the pairwise \emph{distance-variance matrix} of a structural ensemble as
a coherence diagnostic. Because the distance-variance matrix captures
inter-residue coordination rather than coordinate-space spread, such a
measure is complementary to---not a surrogate for---coordinate-covariance
spectral entropy.

In this work, we evaluate the Spectral Coherence Index (SCI), a
model-free, rotation-invariant metric derived from the participation
ratio effective rank of the inter-model pairwise distance-variance
matrix, in an expanded Main110 validation setting designed not to
redefine the metric, but to test whether its coherence signal survives
broader biological heterogeneity and stricter validation regimes than
our earlier internal 27-protein pilot could support. The main
contributions are:

\begin{enumerate}
\item \textbf{Expanded validation beyond the internal pilot.}
      We scale from an internal 27-protein pilot to a canonical
      Main110 cohort of 110 NMR ensembles plus a fixed 11-protein
      holdout, showing that the SCI signal survives broader cohort
      heterogeneity while also revealing softer specificity and
      threshold non-transferability that were not visible at pilot
      scale.

\item \textbf{Canonical metric and grouped inference layer.} We retain the canonical
      definition $\mathrm{SCI}=1-r_\text{eff}/d$, where $r_\text{eff}$
      is the participation ratio of the positive eigenvalue spectrum of
      the distance-variance matrix and $d=\min(L,M{-}1)$ normalizes
      degrees of freedom ($L$: number of residues; $M$: number of
      models). We establish grouped UniProt or analysis-group inference
      as the primary paper layer, with PDB-level analysis reported as a
      sensitivity check.

\item \textbf{Benchmarking, generalization, and normalization analysis.} We
      compare SCI with $\sigma_{R_g}$, SCI+$\sigma_{R_g}$, PCA variance
      ratio, contact density, and a QC-augmented multifeature model
      under 5-fold grouped cross-validation, leave-one-function-class-out
      validation, and independent holdout testing. We further show that
      raw spectral quantities ($r_\text{eff}$, top-1 energy fraction,
      top-3 energy fraction) and calibrated SCI models augmented with
      residue count and model count nearly recover the lost
      discrimination, explaining why Main110 is harder than the pilot.

\item \textbf{Residue-level biological validation.} We validate SCI-derived residue
      contributions against experimental RMSF and Gaussian network model
      predictions across 110 proteins, including paired statistical
      comparison and outlier analysis, and we treat allosteric, apo/holo,
      and intrinsically disordered examples as contextual case studies.

\item \textbf{Practical multimetric QC guidance.} We show that coherence (SCI),
      amplitude ($\sigma_{R_g}$), and eigenvector smoothness are
      complementary diagnostics. Function-class breakdowns, threshold
      error analyses, ensemble-size sensitivity, and synthetic
      failure-mode stress tests support their combined use, while
      making clear that fixed QC thresholds should be interpreted as
      screening heuristics rather than universal pass/fail rules.
\end{enumerate}

The remainder of this paper is organized as follows.
Section~\ref{sec:related} reviews related work on protein ensemble
analysis and spectral methods. Section~\ref{sec:methods} describes the
dataset, the SCI formulation, baseline metrics, and statistical
procedures. Section~\ref{sec:results} presents the main experimental
results. Section~\ref{sec:discussion} discusses methodological
implications, limitations, and practical guidance.
Section~\ref{sec:conclusion} concludes.

\section{Related Work}
\label{sec:related}

\subsection{Protein Ensembles: Experiment, Prediction, and Quality}

The energy landscape theory established that proteins exist as
statistical ensembles of conformational
states~\cite{frauenfelder1991energy, dill2012protein}, with allosteric
regulation, catalysis, and molecular recognition depending on
coordinated structural fluctuations~\cite{henzler2007dynamic,
boehr2009role, smock2010interdomain}. NMR spectroscopy provides direct
experimental access to these ensembles through multi-model PDB
depositions~\cite{bonvin1995nmr, clore2004accurate,
montelione2013recommendations}, capturing coherent motions such as
hinge bending and domain closure~\cite{gerstein1994structural,
hayward1998systematic}. On the computational side,
AlphaFold2~\cite{jumper2021highly},
RoseTTAFold~\cite{baek2021accurate},
ESMFold~\cite{lin2023evolutionary}, and protein language
models~\cite{rives2021biological} have achieved high single-structure
accuracy, spurring ensemble extensions via sequence
subsampling~\cite{wayment2024predicting}, stochastic
sampling~\cite{del2022sampling}, diffusion
models~\cite{jing2024alphafold}, and trainable
reimplementations~\cite{ahdritz2024openfold}. Molecular dynamics
remains the physics-based gold standard for conformational
sampling~\cite{shaw2010atomic, hollingsworth2018molecular} but is
computationally expensive. This proliferation of ensemble sources
creates an urgent need for quality metrics that assess whether the
resulting variation is structurally coherent or merely noise-like. Existing geometric
metrics---pairwise RMSD~\cite{diamond1992relationship,
kufareva2012rmsd}, radius of gyration
statistics~\cite{lobanov2008radius}, and contact map
variability~\cite{vendruscolo2002three}---each have well-known
limitations (Section~\ref{sec:introduction}). Statistical approaches such as structure
clustering~\cite{li2006cd}, backbone entropy~\cite{best2006relation},
TM-score~\cite{zhang2004scoring}, and ensemble validation tools
(PSVS~\cite{bhattacharya2007evaluating},
wwPDB-NMR~\cite{montelione2013recommendations}) focus on restraint
satisfaction or single-structure quality rather than spectral
organization of conformational variation.

\subsection{Spectral Methods and Effective Rank}

Spectral decomposition has a rich history in structural biology. The
Gaussian Network Model~\cite{bahar1997direct, haliloglu1997gaussian}
and the Anisotropic Network Model~\cite{atilgan2001anisotropy}
decompose connectivity matrices into normal modes, predicting
B-factors~\cite{yang2006ognm}, identifying functional
sites~\cite{yang2005catalytic}, and revealing allosteric
pathways~\cite{chennubhotla2007signal}. Quasi-harmonic
analysis~\cite{garcia1992large} and essential
dynamics~\cite{amadei1993essential, david2014principal, balsera1996principal}
apply PCA to Cartesian trajectories but require structural
superposition, introducing rotation-frame artifacts.
Distance-based representations such as multidimensional
scaling~\cite{borg2005modern} avoid this limitation.
Scheurer et al.~\cite{scheurer2015exploration} applied spectral entropy---the
Shannon-entropy effective rank of coordinate-covariance
eigenspectra---to assess \emph{sampling completeness} in molecular dynamics
trajectories. The participation ratio,
originally introduced to characterize eigenstate
localization~\cite{bell1970atomic}, quantifies how many components of a
distribution contribute meaningfully: for a normalized spectrum
$\{\tau_k\}$, $r_\text{eff} = 1/\sum_k \tau_k^2$ equals~1 when all
weight is on a single component and equals the number of
components~$n^+$ when weight is uniform. Roy and Vetterli~\cite{roy2007effective} formalized
effective rank using Shannon entropy of singular values; the participation
ratio is a mathematically related alternative~\cite{vershynin2018high},
and in practice the two formulas produce near-identical rankings
(Section~\ref{sec:res_reff}, $r > 0.95$).
The key distinction between SCI and prior spectral-entropy approaches
therefore lies not in the effective-rank formula but in the
\emph{input representation}: the pairwise distance-variance matrix.
This makes SCI rotation-invariant and diagnostic of inter-residue
coordination, whereas coordinate-covariance spectral entropy measures
sampling breadth.

\subsection{Ensemble Quality in Biomedical Informatics}

Structural ensemble quality has direct implications for downstream
health informatics tasks. Ensemble docking methods for binding-site
prediction~\cite{amaro2018ensemble} can propagate artifacts from
incoherent ensembles. Epitope mapping leveraging conformational
diversity~\cite{thornton1986antigenic} and variant effect prediction
tools incorporating structural dynamics~\cite{ponzoni2018structural}
both depend on ensemble quality as a prerequisite for reliable
inference. More broadly, automated quality gates for structural
data---analogous to QC metrics in genomics
pipelines~\cite{ewels2016multiqc}---are essential for maintaining
reproducibility in computational analyses that inform clinical
decision-making.

\section{Methods}
\label{sec:methods}

\subsection{Dataset Construction}
\label{sec:dataset}

\subsubsection{Expanded Main110 Cohort}
We promoted an expanded cohort of $N = 110$ NMR multi-model protein
structures from a 125-candidate screen of PDB NMR entries, requiring
usable C$\alpha$ coordinates, successful preprocessing, and resolved
metadata after UniProt-mapping quality control. Four non-holdout
candidates were excluded because of unresolved mapping or hard QC
failures, and one mapped unresolved case (6LF5) was retained as a
PDB-level fallback analysis group. The final main cohort spans proteins
of 30--403 residues with 10--30 models per entry (median
$M = 20$) across 11 broad function classes. C$\alpha$ coordinates were
extracted from mmCIF files using Biotite.

\subsubsection{Independent Holdout}
An additional 11 NMR proteins were reserved as a fixed independent
holdout and were not used to define the canonical main-cohort summary
statistics or grouped cross-validation models. The holdout spans
27--75 residues with 10--25 models per entry (median $M = 20$) and is
used as an external NMR-vs-synthetic-incoherent-control validation set.

\subsubsection{Synthetic Incoherent Controls}
As a negative control representing ensembles with incoherent structural
variation, we generated synthetic incoherent controls by perturbing the first NMR
conformer along $K = 2$ randomly sampled global deformation modes.
Specifically, for each protein with reference coordinates
$\mathbf{x}_0 \in \mathbb{R}^{3L}$, the $m$-th synthetic model is
\begin{equation}
\label{eq:synthetic_control}
\mathbf{x}_m = \mathbf{x}_0 + \sum_{k=1}^{K} \alpha_{mk} \, \mathbf{v}_k\,,
\end{equation}
where $\{\mathbf{v}_k\}_{k=1}^{K}$ are orthonormal vectors in
$\mathbb{R}^{3L}$ obtained by Gram--Schmidt orthogonalization of
standard Gaussian random vectors, and
$\alpha_{mk} \sim \mathcal{N}(0, \sigma^2)$ with $\sigma = 0.2$~\AA\
(chosen to match the order of magnitude of NMR C$\alpha$ coordinate
variation; see Section~\ref{sec:meth_failure} for a discussion of
perturbation magnitudes). $K = 2$ modes ensure that the control is
low-rank yet minimally coherent in distance-variance space; higher~$K$
would produce a more diffuse control (tested as a separate failure mode
with $K = 10$). Each synthetic ensemble has the same number of models
$M$ as the corresponding NMR ensemble in both the main and holdout
splits. Although the perturbation is
low-rank in coordinate space ($K = 2$), the resulting patterns in distance variance
space are typically more diffuse due to the nonlinear distance mapping
(see Section~\ref{sec:results}).

\subsubsection{AlphaFold Single-Structure Reference}
As a reference-only third group, we obtained AlphaFold2-predicted
structures for 40 proteins in the main cohort with resolved mappings and
usable AlphaFold coverage from the AlphaFold Protein Structure
Database~\cite{jumper2021highly}. Each prediction provides a single
model ($M = 1$), precluding ensemble-based SCI computation. We
therefore employ a single-structure spectral proxy
(Section~\ref{sec:proxy}). Per-residue confidence scores (pLDDT) were
extracted from B-factor fields. Sequence matching required a
protein-level metadata mapping and compatible chain coverage; these
AlphaFold results are reported only as a reference analysis and are not
directly comparable to the ensemble-based SCI values.

\begin{figure*}[!t]
\centering
\includegraphics[width=\textwidth]{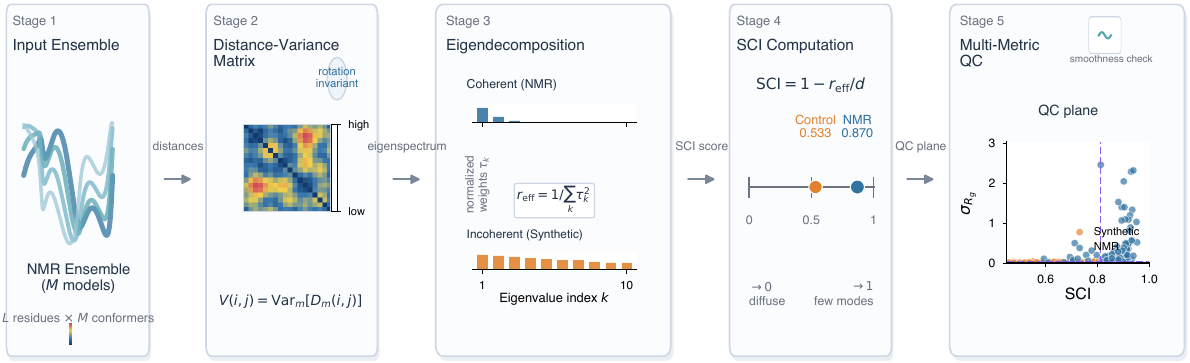}
\caption{Conceptual overview of the SCI workflow. Starting from an NMR
ensemble of $M$ conformers, pairwise distance variation is summarized in
the distance-variance matrix $V$, eigendecomposed to obtain the
normalized positive-spectrum weights $\tau_k$, compressed into the
bounded score $\mathrm{SCI}=1-r_\mathrm{eff}/d$, and finally interpreted
jointly with $\sigma_{R_g}$ in a multimetric QC plane. The SCI anchors
shown in Stage~4 correspond to the current Main110 mean SCI for NMR
ensembles (0.870) and matched synthetic incoherent controls (0.533), and
the QC plane in Stage~5 uses the current Main110 data with the manuscript's
screening thresholds.}
\label{fig:pipeline}
\end{figure*}

Figure~\ref{fig:pipeline} summarizes the SCI workflow from ensemble
input through spectral compression to multimetric QC interpretation.

\subsection{Spectral Coherence Index (SCI)}
\label{sec:sci}

Throughout this study, we define the positive class as NMR ensembles and
use SCI directly as the classification score: higher SCI indicates more
coherent variation, corresponding to the NMR ensemble class.

\subsubsection{Distance Variance Matrix}
Consider a structural ensemble with $L$ residues per model and $M$
conformers. For the $m$-th model, define the pairwise
C$\alpha$--C$\alpha$ distance matrix
\begin{equation}
\label{eq:distmat}
D_m(i,j) = \|\mathbf{r}_m^{(i)} - \mathbf{r}_m^{(j)}\|_2\,,
\quad 1 \leq i, j \leq L\,,
\end{equation}
where $\mathbf{r}_m^{(i)} \in \mathbb{R}^3$ is the coordinate of the
$i$-th C$\alpha$ atom in model~$m$. The \emph{distance-variance matrix}
$V \in \mathbb{R}^{L \times L}$ is defined entry-wise as
\begin{equation}
\label{eq:varmat}
V(i,j) = \mathrm{Var}_m\bigl[D_m(i,j)\bigr]
= \frac{1}{M} \sum_{m=1}^{M}
\bigl(D_m(i,j) - \bar{D}(i,j)\bigr)^2\,,
\end{equation}
where $\bar{D}(i,j) = \frac{1}{M}\sum_{m=1}^{M} D_m(i,j)$ is the
mean distance. We use the population variance ($1/M$); using the
sample variance ($1/(M{-}1)$) would yield an identical SCI, since the
constant factor cancels in the normalized spectrum~$\tau_k$. Each entry
$V(i,j) \geq 0$ is the scalar variance of the $(i,j)$~pairwise
distance across $M$~models.

Several properties of $V$ are important:
\begin{itemize}
\item $V$ is symmetric with non-negative entries, but is \emph{not}
      guaranteed positive semi-definite (PSD). It is an entry-wise
      variance map, not a covariance matrix in the sense of
      $\mathrm{Cov}(\mathrm{vec}(D), \mathrm{vec}(D)^\top)$.

\item The diagonal entries satisfy $V(i,i) = \mathrm{Var}_m[D_m(i,i)]
      = \mathrm{Var}_m[0] = 0$ for all~$i$, since self-distances are
      identically zero. Therefore $\mathrm{tr}(V) = \sum_i V(i,i) = 0$,
      which implies $\sum_k \lambda_k = 0$: the positive and negative
      eigenvalue masses are exactly equal. The retained
      positive-spectrum fraction is thus $0.50$ for every ensemble---a
      structural identity, not a numerical artifact.

\item Because $D_m(i,j)$ depends only on inter-atomic distances, $V$ is
      \emph{rotation-invariant}: no structural superposition or
      reference frame is required.

\item Although $V$ is not PSD, its positive eigenspectrum provides a
      compact summary of coordinated variance patterns across residue
      pairs. We define SCI on the positive part of the spectrum only;
      negative eigenvalues are discarded. The fraction of retained
      spectral mass is reported in the project outputs as a transparency
      diagnostic.
\end{itemize}

\subsubsection{Eigendecomposition and Normalized Spectrum}
Let $\lambda_1 \geq \lambda_2 \geq \cdots \geq \lambda_L$ be the
eigenvalues of $V$. Since $V$ is not PSD, some $\lambda_k$ may be
negative. We retain only the positive eigenvalues exceeding a numerical
threshold~$\varepsilon = 10^{-10}$:
\begin{equation}
\label{eq:filter}
\Lambda^{+} = \{\lambda_k : \lambda_k > \varepsilon\}\,,
\quad n^{+} = |\Lambda^{+}|\,.
\end{equation}
Denote the filtered eigenvalues as
$\lambda_1^{+} \geq \cdots \geq \lambda_{n^{+}}^{+}$. The normalized
spectrum is
\begin{equation}
\label{eq:tau}
\tau_k = \frac{\lambda_k^{+}}{\displaystyle\sum_{j=1}^{n^{+}} \lambda_j^{+}}\,,
\quad k = 1, \ldots, n^{+}\,,
\end{equation}
satisfying $\tau_k \geq 0$ and $\sum_k \tau_k = 1$, forming a
discrete probability distribution over variance modes.

\subsubsection{Effective Rank via Participation Ratio}
The participation ratio effective rank~\cite{roy2007effective,
vershynin2018high} of the normalized spectrum is
\begin{equation}
\label{eq:reff}
r_\text{eff}
= \frac{\bigl(\sum_{k} \tau_k\bigr)^2}{\sum_{k} \tau_k^2}
= \frac{1}{\sum_{k} \tau_k^2}\,.
\end{equation}
This quantity has the following extremal behavior:
\begin{itemize}
\item $r_\text{eff} = 1$ when all variance is concentrated in a single
      mode ($\tau_1 = 1$);
\item $r_\text{eff} = n^{+}$ when variance is uniformly distributed
      ($\tau_k = 1/n^{+}$ for all~$k$).
\end{itemize}
Thus $r_\text{eff}$ can be interpreted as the effective number of
independent variation modes.

\subsubsection{SCI Definition}
The Spectral Coherence Index is defined as
\begin{equation}
\label{eq:sci}
\mathrm{SCI} = 1 - \frac{r_\text{eff}}{d}\,,
\quad d = \min(L,\; M{-}1)\,.
\end{equation}

The normalization constant $d = \min(L, M{-}1)$ serves as an effective
degrees-of-freedom scale. The $M{-}1$ term reflects the fact that
$M$~ensemble members yield $M{-}1$ independent deviations from the
mean, analogous to the degrees-of-freedom correction in sample variance.
We note that $V$ is not a sample covariance matrix \emph{sensu stricto},
so $M{-}1$ is not a strict rank upper bound: empirically, $n^{+}$
(mean~$\approx 43$ for $L \approx 90$, $M \approx 20$) exceeds $M{-}1$
because the entry-wise variance map is not generated by a
rank-$(M{-}1)$ process. However, $\min(L, M{-}1)$ empirically yields
$\mathrm{SCI} \in [0,1]$ for all ensembles in our dataset and scales
appropriately with both protein size and ensemble size; the bound is
not a mathematical guarantee, since a perfectly uniform positive
spectrum could in principle produce $r_\text{eff} > d$. Empirical
validation in the project outputs confirms that all
$r_\text{eff}$ values fall well below~$d$.

SCI has the following interpretation:
\begin{itemize}
\item $\mathrm{SCI} \to 1$: the spectrum is highly concentrated;
      distance variation across models is dominated by few coherent
      modes, consistent with coordinated structural deformations (e.g.,
      hinge bending, domain closure).
\item $\mathrm{SCI} \to 0$: the spectrum is diffuse; distance variation
      is spread across many modes, indicating incoherent, noise-like
      fluctuations.
\end{itemize}
Crucially, SCI measures the \emph{coherence} of variation---the degree
to which pairwise distance changes are coordinated across residue
pairs---not its magnitude.

SCI belongs to the broader family of effective-rank measures, which
includes the Shannon-entropy variant used by Scheurer et
al.~\cite{scheurer2015exploration} to assess conformational sampling
completeness from coordinate-covariance eigenspectra. The choice of
participation ratio versus Shannon entropy is a minor technical detail:
both yield near-identical SCI values (Section~\ref{sec:res_reff},
$r > 0.95$). The substantive distinction is the input representation:
SCI summarizes the eigenspectrum of the distance-variance matrix~$V$,
making it rotation-invariant and diagnostic of inter-residue
coordination, whereas spectral entropy of coordinate covariance quantifies
how broadly conformational space has been explored. The two measures
answer complementary questions about ensemble quality.

We additionally define the \emph{spectral dispersion index}
$\mathrm{DI} = 1 - \mathrm{SCI} = r_\text{eff}/d$, which increases
with spectral diffuseness. Higher DI does not imply greater conformational
diversity; it indicates that distance changes across models lack
coordinated structure.

\subsubsection{Single-Structure Spectral Proxy}
\label{sec:proxy}
For AlphaFold single structures ($M = 1$), the distance-variance matrix
is identically zero. As an alternative, we compute a spectral proxy from
the double-centered Gram matrix of the pairwise distance
matrix~\cite{borg2005modern}:
\begin{equation}
\label{eq:gram}
B = -\tfrac{1}{2}\, H\, D^{\circ 2}\, H\,,
\quad H = I_L - \frac{1}{L}\mathbf{1}_L\mathbf{1}_L^\top\,,
\end{equation}
where $D^{\circ 2}$ denotes the element-wise squared distance matrix and
$H$ is the centering matrix. The SCI formula
\eqref{eq:reff}--\eqref{eq:sci} is then applied to $B$ with $d = L$.
This proxy measures intrinsic fold geometry spectral concentration
rather than cross-model variation coherence, and is \emph{not} directly
comparable to the ensemble-based SCI ($d = \min(L, M{-}1)$). All
cross-group comparisons involving AlphaFold are therefore interpretive.

\subsection{Baseline Metrics}
\label{sec:baselines}

To contextualize SCI, we compare it against three baseline metrics
computable from the same C$\alpha$ ensembles:

\begin{enumerate}
\item \textbf{Radius of gyration standard deviation}
      ($\sigma_{R_g}$): the standard deviation of the radius of gyration
      across ensemble members, where
      $R_g(S_m) = \sqrt{\frac{1}{L}\sum_{i=1}^{L}
      \|\mathbf{r}_m^{(i)} - \bar{\mathbf{r}}_m\|^2}$
      and $\bar{\mathbf{r}}_m$ is the centroid of model~$m$. This
      captures global size fluctuations but is insensitive to local
      conformational changes.

\item \textbf{PCA variance ratio} ($\lambda_1^\text{PCA}/\Sigma$): the
      fraction of total variance explained by the first principal
      component, computed from the $3L$-dimensional superposed
      coordinate vectors. This requires structural alignment and is
      sensitive to reference frame choice.

\item \textbf{Contact density}: the mean number of C$\alpha$--C$\alpha$
      contacts per residue (threshold~8~\AA), averaged across models.
      This metric is not expected to discriminate NMR from synthetic
      controls and serves as a negative baseline.
\end{enumerate}

\subsection{Statistical Analysis}
\label{sec:stats}

All primary discrimination analyses treat experimental NMR ensembles as
the positive class and matched synthetic incoherent controls as the
negative class.

\subsubsection{Primary and Sensitivity Analyses}
The canonical manuscript endpoint is a grouped main-cohort comparison in
which NMR and matched synthetic incoherent controls are paired by
UniProt-derived analysis group. The main cohort contains 110 analysis
groups: 109 resolved UniProt groups and one retained PDB-level fallback
(6LF5). We report AUC-ROC, Cliff's delta, paired permutation tests, and
bootstrap confidence intervals for the mean paired SCI difference. A
PDB-level analysis on the same 110 proteins is reported as a sensitivity
check, and the fixed 11-protein holdout is evaluated independently as an
external NMR-vs-synthetic-incoherent-control test set. Legacy27 results
are retained only as historical context and not as a separate inferential
cohort.

\subsubsection{Reference Comparison and Thresholding}
For the reference-only AlphaFold analysis, a Kruskal--Wallis test
assesses differences across three groups (NMR, synthetic incoherent controls, and
AlphaFold proxy), followed by pairwise Mann--Whitney~$U$ tests with
multiple-comparison correction. Because the AlphaFold proxy uses the
single-structure Gram-matrix formulation of Section~\ref{sec:proxy},
these comparisons are interpretive rather than directly comparable to
the ensemble-based SCI results. We estimate operating thresholds with
Youden's $J$ and summarize threshold uncertainty by bootstrap
confidence intervals.

\subsubsection{Generalization Analyses}
Generalization is evaluated at three levels. First, all manuscript
results use 5-fold grouped cross-validation with
\texttt{StratifiedGroupKFold(n\_splits=5, shuffle=True, random\_state=42)},
grouping by \texttt{analysis\_group\_id} and stratifying by the binary
NMR/control label. The code retains \texttt{GroupKFold} as a
backward-compatible fallback for older environments, but it was not
used in the reported results. This preserves paired NMR/control samples
within analysis-group splits on the main cohort. Second,
leave-one-function-class-out validation tests transfer across the 11
annotated protein function classes. Third, the independent holdout is
scored using models fitted on the main cohort only. These analyses are
reported for SCI, $\sigma_{R_g}$, SCI+$\sigma_{R_g}$, and a QC-full
model that combines SCI, $\sigma_{R_g}$, PCA variance ratio, and
top-eigenvector smoothness. To investigate why SCI softens on Main110,
we repeat the same grouped-CV, leave-one-function-class-out, and holdout
regimes for raw spectral quantities ($r_\text{eff}$, top-1 energy
fraction, top-3 energy fraction) and for calibrated SCI models that add
residue count and/or model count as covariates.

\subsubsection{Biological Validation}
We summarize three biological validation layers. At the residue level,
Spearman correlations compare SCI-derived residue contributions with
experimental RMSF and with GNM-predicted RMSF, and paired Wilcoxon tests
compare these two per-protein correlation distributions. We additionally
report negative and weak-correlation outliers. At the protein level, we
evaluate curated allosteric and apo/holo pairs to determine whether SCI
tracks coherence differently from amplitude-sensitive metrics, but we
interpret these small-sample analyses as contextual case studies rather
than formal validation. Finally, available intrinsically disordered
proteins are summarized separately as high-flexibility examples.

\subsubsection{Eigenvector Smoothness Score}
Let $\mathbf{u}_1$ denote the top eigenvector of the distance-variance
matrix $V$. To assess whether this dominant mode respects the protein's
chain topology, we define a smoothness score via the Rayleigh quotient
on the sequential residue graph Laplacian $\mathcal{L}$. The graph is
the path graph on $L$ residues (adjacency $A_{i,i+1} = 1$), and
$\mathcal{L} = D_\mathrm{deg} - A$ is the combinatorial Laplacian. We
compute $E = \mathbf{u}_1^\top \mathcal{L}\, \mathbf{u}_1 /
\mathbf{u}_1^\top \mathbf{u}_1$ and define smoothness $= 1/(1+E)
\in [0,1]$, where values near~1 indicate that the dominant mode varies
slowly along the chain (spatially coherent), and lower values indicate
rapid residue-to-residue sign changes (spatially incoherent).

\subsubsection{Robustness and Failure-Mode Diagnostics}
\label{sec:meth_failure}
To stress-test the multimetric QC framework, we generate four synthetic
failure types from the reference NMR ensembles used by the QC stress-test
pipeline:
\begin{enumerate}
\item \textbf{i.i.d.\ Gaussian noise.} Per-atom i.i.d.\ perturbations
      $\mathcal{N}(0, 0.2^2)$~\AA\ are added independently to each
      coordinate of each model.
\item \textbf{Generator mode collapse.} Three conformations are drawn
      uniformly at random from the NMR ensemble and replicated to fill
      $M$~models, with per-atom jitter $\mathcal{N}(0, 0.01^2)$~\AA\
      added to each copy.
\item \textbf{Residue shuffle.} Each residue in each model receives an
      independent random displacement of magnitude
      $|\mathcal{N}(0, 0.2)|$~\AA\ in a uniformly random direction,
      destroying spatial correlation between neighboring residues.
\item \textbf{High-rank perturbation ($K = 10$).} Ten randomly sampled
      global deformation modes with $\sigma = 0.1$~\AA\ are applied
      (using Eq.~\eqref{eq:synthetic_control} with $K = 10$), producing
      spectrally diffuse variation.
\end{enumerate}
The perturbation magnitudes ($\sigma = 0.2$~\AA\ for i.i.d.\ and
shuffle, $\sigma = 0.1$~\AA\ for high-rank) are chosen to match the
order of magnitude of coordinate-level NMR ensemble variation (typical
C$\alpha$ RMSD $\approx 0.5$--$2.0$~\AA) while keeping the artifacts
subtle enough that naive metrics may miss them. The $K = 2$ baseline
control (Section~\ref{sec:dataset}) uses the same low-rank perturbation
scheme but with fewer modes, producing a clear coherence deficit.

\begin{table*}[!t]
\caption{Expanded cohort summary used in this manuscript. The grouped
primary endpoint comprises 110 analysis groups: 109 UniProt-mapped
groups and one retained PDB-level fallback group.}
\label{tab:dataset}
\centering
\begin{tabular}{lcccc}
\hline
Cohort & Role & $n$ & Residues & Models per entry \\
\hline
Main110 NMR & grouped primary and PDB sensitivity & 110 & 30--403 & 10--30 (median 20) \\
Holdout NMR & independent external validation & 11 & 27--75 & 10--25 (median 20) \\
AlphaFold proxy & reference-only single-structure analysis & 40 & matched main-cohort subset & 1 \\
Excluded candidates & unresolved mapping or hard QC failure & 4 & --- & --- \\
\hline
\end{tabular}
\end{table*}

\section{Results}
\label{sec:results}

\begin{figure*}[!t]
\centering
\includegraphics[width=\textwidth]{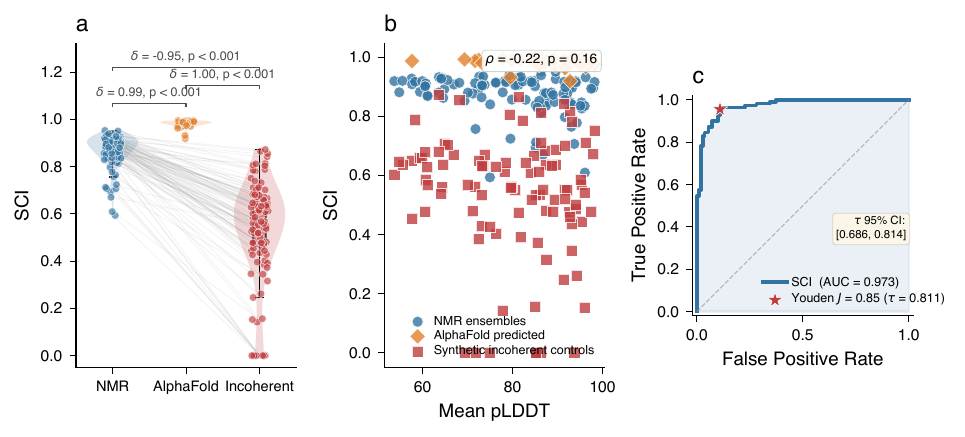}
\caption{Main results. \textbf{(a)}~SCI distribution across the Main110
NMR cohort, matched synthetic incoherent controls, and 40 AlphaFold
single-structure proxies. Brackets indicate pairwise Cliff's $\delta$
and corrected $p$-values. \textbf{(b)}~SCI versus mean pLDDT for the
AlphaFold reference structures, showing a weak non-significant
association. \textbf{(c)}~ROC curve for the grouped primary Main110
task (AUC~$= 0.973$); the star marks the Youden-$J$ operating point
($\tau = 0.811$).}
\label{fig:main}
\end{figure*}

\subsection{SCI Distinguishes NMR Ensembles from Synthetic Incoherent Controls}
\label{sec:res_binary}

The grouped primary Main110 analysis shows a clear separation between
experimental NMR ensembles and matched synthetic incoherent controls.
Mean SCI was $0.870 \pm 0.068$ for NMR versus $0.533 \pm 0.209$ for the
matched synthetic incoherent controls, with a mean paired difference of $0.337$
(95\% bootstrap CI: $[0.305, 0.370]$) and paired permutation
$p < 10^{-4}$. The effect size remained very large
(Cliff's $\delta = -0.945$), indicating that higher SCI is strongly
associated with the experimental ensemble class. Using SCI directly as
the score yielded grouped AUC-ROC $= 0.9726$
(Fig.~\ref{fig:main}c). The Youden-$J$ operating point was
$\tau = 0.811$, corresponding to 95.5\% sensitivity and 89.1\%
specificity; the bootstrap confidence interval for the threshold was
$[0.686, 0.814]$.

The grouped result is the manuscript's primary inferential layer because
the expanded cohort is analyzed at the resolved UniProt or
analysis-group level rather than by counting every archival entry as an
independent biological target. In the canonical Main110 set, 109 groups
map cleanly to UniProt identifiers and one group is retained through
the fallback identifier \texttt{PDB:6LF5}; this bookkeeping avoids
pseudo-replication without discarding a structurally valid ensemble.

The same conclusion persisted in the two confirmatory layers used in
this manuscript. A PDB-level sensitivity analysis remained nearly
unchanged (AUC $= 0.9716$, Cliff's $\delta = -0.943$,
$\tau = 0.813$), showing that the coherence signal is not created by
the grouping rule. The fixed 11-protein holdout also preserved strong
separation, with mean SCI $0.9068 \pm 0.0290$ for NMR versus
$0.7354 \pm 0.0865$ for matched synthetic incoherent controls, AUC $= 0.9835$, and a
descriptive holdout threshold $\tau = 0.866$ achieving 100\%
sensitivity and 90.9\% specificity. The holdout controls are therefore
more challenging than the main synthetic incoherent controls in absolute SCI, yet
the ranking signal remains robust.

Relative to the internal 27-protein pilot rerun, however, the expanded
benchmark is modestly but meaningfully harder. The pilot achieved
AUC $= 0.9877$, Cliff's $\delta = -0.9753$, and a Youden threshold
$\tau = 0.8249$ with 100\% sensitivity and 92.6\% specificity. Main110
therefore reduces AUC by 0.0151 and specificity by 3.5 percentage
points while retaining strong sensitivity. This softening is
informative rather than discouraging: the broader cohort spans
11 function classes and a wider range of residue counts and model
counts, exposing heterogeneity that the pilot could not resolve.

Error analysis makes the failure pattern explicit. At the grouped
threshold, 12 of 110 NMR groups fall below $\tau$ and 6 of 110
synthetic incoherent controls rise above it. These cases are not evenly
distributed. Low-model ensembles ($M \leq 10$) represent only 11 of 110
groups (10.0\%) but account for 10 of the 12 NMR misses (83.3\%), an
approximately 8.3-fold enrichment. Short proteins ($L \leq 50$)
represent 13 of 110 groups (11.8\%) but account for 5 of the 6 control
overcalls (83.3\%), an approximately 7.1-fold enrichment. Thus Main110
reveals two distinct SCI stress points: NMR misses are disproportionately
concentrated among low-model ensembles, whereas control overcalls are
enriched among very short proteins.

Function-class summaries reinforce the same interpretation
(Table~\ref{tab:function_breakdown}). By mean SCI, chaperone, other,
toxin, enzyme, and transport proteins are the weakest classes, and the
viral split is the hardest leave-one-function-class-out transfer
setting for SCI (AUC $= 0.906$). In contrast, $\sigma_{R_g}$ remains at
or above 0.938 AUC on every held-out function class. Representative
low-margin and threshold-failed groups are listed in
Table~\ref{tab:hard_cases}.

\begin{table*}[!t]
\caption{Statistical summary for the Main110 manuscript.}
\label{tab:stats}
\centering
\begin{tabular}{llrl}
\hline
Test & Statistic & Value & Note \\
\hline
\multicolumn{4}{l}{\emph{Primary discrimination: grouped Main110}} \\
Paired permutation & $p$ & $< 10^{-4}$ & $n = 10{,}000$ permutations \\
AUC-ROC & --- & 0.9726 & SCI as score \\
Cliff's $\delta$ & $\delta$ & $-0.945$ & higher SCI in NMR ensembles \\
Paired difference & mean & 0.337 & 95\% CI: $[0.305, 0.370]$ \\
Youden's $J$ & $\tau^*$ & 0.811 & sens.\ 95.5\%, spec.\ 89.1\% \\
Threshold 95\% CI & --- & $[0.686, 0.814]$ & bootstrap \\[4pt]
\multicolumn{4}{l}{\emph{Sensitivity and holdout confirmation}} \\
PDB-level sensitivity & AUC-ROC & 0.9716 & $\delta = -0.943$, $\tau^* = 0.813$ \\
Independent holdout & AUC-ROC & 0.9835 & $\delta = -0.967$, $\tau^* = 0.866$ \\[4pt]
\multicolumn{4}{l}{\emph{Reference-only three-group comparison}} \\
Kruskal--Wallis & $H$ & 205.52 & $p = 2.36 \times 10^{-45}$ \\
Mann--Whitney (NMR vs.\ incoherent control) & $U$ & 11769.0 & $p_\mathrm{corr} = 2.67 \times 10^{-33}$ \\
Mann--Whitney (NMR vs.\ AF) & $U$ & 29.0 & $p_\mathrm{corr} = 8.56 \times 10^{-20}$ \\
Mann--Whitney (control vs.\ AF) & $U$ & 0.0 & $p_\mathrm{corr} = 2.67 \times 10^{-20}$ \\
\hline
\end{tabular}
\end{table*}

\begin{table*}[!t]
\caption{Function-class summary for the grouped Main110 cohort. Mean
SCI and mean $\sigma_{R_g}$ are reported for the NMR side of each
analysis group, together with leave-one-function-class-out (LOFO) AUC
for SCI and $\sigma_{R_g}$ and the number of threshold-failed NMR groups
and threshold-exceeding synthetic incoherent controls.}
\label{tab:function_breakdown}
\centering
\begin{tabular}{lccccccc}
\hline
Function class & $n$ groups & Mean SCI & Mean $\sigma_{R_g}$ (\AA) & SCI LOFO & $\sigma_{R_g}$ LOFO & NMR $< \tau$ & Control $> \tau$ \\
\hline
chaperone & 3 & 0.824 & 0.292 & 1.000 & 1.000 & 1 & 0 \\
other & 6 & 0.830 & 0.390 & 0.944 & 1.000 & 1 & 1 \\
toxin & 5 & 0.843 & 0.106 & 0.960 & 1.000 & 1 & 0 \\
enzyme & 22 & 0.850 & 0.484 & 0.983 & 0.977 & 3 & 0 \\
transport & 13 & 0.854 & 0.624 & 0.964 & 0.988 & 3 & 0 \\
viral & 8 & 0.858 & 0.189 & 0.906 & 0.984 & 1 & 1 \\
immune & 9 & 0.870 & 0.206 & 0.963 & 0.938 & 1 & 2 \\
signaling & 15 & 0.889 & 0.441 & 0.996 & 0.987 & 1 & 1 \\
idp & 7 & 0.894 & 0.393 & 0.959 & 0.980 & 0 & 1 \\
structural & 9 & 0.899 & 2.188 & 1.000 & 1.000 & 0 & 0 \\
dna\_rna\_binding & 13 & 0.911 & 1.472 & 1.000 & 1.000 & 0 & 0 \\
\hline
\end{tabular}
\end{table*}

\begin{table*}[!t]
\caption{Representative low-margin and threshold-sensitive Main110
groups. $\Delta$SCI denotes the diverse-minus-control SCI margin within
each matched group.}
\label{tab:hard_cases}
\centering
\begin{tabular}{lccccc}
\hline
PDB & Function class & $L$ & $M$ & $\Delta$SCI & Error pattern \\
\hline
2KUX & other & 30 & 20 & 0.033 & control above $\tau$, $L \leq 50$ \\
2MCR & immune & 36 & 30 & 0.059 & control above $\tau$, $L \leq 50$ \\
1SKH & idp & 30 & 22 & 0.060 & control above $\tau$, $L \leq 50$ \\
7OFO & viral & 30 & 20 & 0.066 & control above $\tau$, $L \leq 50$ \\
5LBJ & signaling & 35 & 20 & 0.075 & control above $\tau$, $L \leq 50$ \\
6QES & immune & 40 & 10 & 0.118 & NMR below $\tau$, $M \leq 10$, $L \leq 50$ \\
2LM3 & enzyme & 205 & 10 & 0.694 & NMR below $\tau$, $M \leq 10$ \\
1Q5F & other & 156 & 10 & 0.593 & NMR below $\tau$, $M \leq 10$ \\
\hline
\end{tabular}
\end{table*}

\subsection{Reference Analysis: AlphaFold Single-Structure Proxy}
\label{sec:res_three}

The AlphaFold comparison remains methodologically separate from the main
ensemble analysis. Because each AlphaFold prediction comprises a single
model, SCI is computed via the Gram-matrix proxy of
Section~\ref{sec:proxy}, which summarizes intrinsic fold-geometry
concentration rather than cross-model coherence. The proxy therefore
serves as a reference analysis, not as a directly comparable third arm
of the primary discrimination task. The AlphaFold proxy values were
high ($0.9813 \pm 0.0155$ for 40 matched structures), a
Kruskal--Wallis test across NMR, synthetic incoherent controls, and
AlphaFold proxies remained highly significant ($H = 205.52$,
$p = 2.36 \times 10^{-45}$), and all pairwise comparisons survived
correction. However, the proxy showed only a weak, non-significant
association with mean pLDDT (Spearman $\rho = -0.224$, $p = 0.165$).
We therefore retain AlphaFold only as a sanity check on the spectral
machinery and not as mechanistic evidence about prediction confidence.

\subsection{Effective Rank and Eigenvalue Diagnostics}
\label{sec:res_reff}

\begin{figure*}[!t]
\centering
\includegraphics[width=\textwidth]{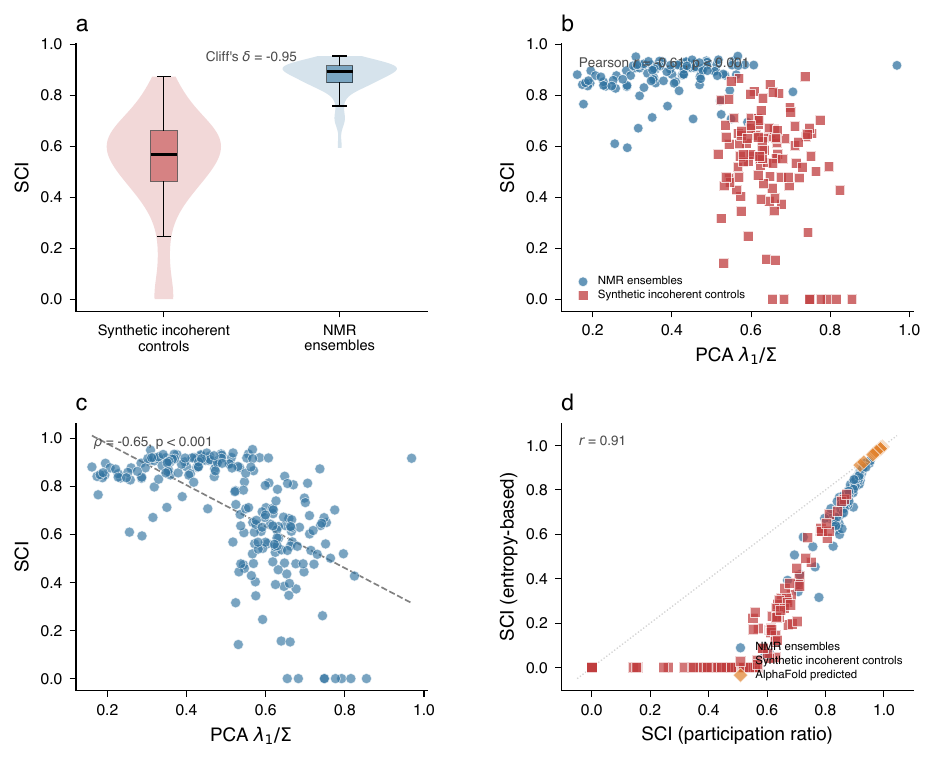}
\caption{Spectral diagnostics. \textbf{(a)}~SCI distribution for
Main110 NMR ensembles versus matched synthetic incoherent controls.
\textbf{(b)}~SCI versus PCA first-component variance ratio.
\textbf{(c)}~Rank-correlation between SCI and PCA variance ratio.
\textbf{(d)}~Participation-ratio SCI versus entropy-based SCI,
illustrating that the main ranking is driven by the distance-variance
representation rather than the specific effective-rank formula.}
\label{fig:spectral}
\end{figure*}

Table~\ref{tab:reff} and Fig.~\ref{fig:spectral} show that the expanded
cohort preserves the same spectral signature seen in the pilot study:
experimental ensembles have much lower effective rank and much more
concentrated eigenspectra than synthetic incoherent controls. Main110 NMR
ensembles had mean $r_\text{eff} = 2.13 \pm 0.67$ versus
$7.93 \pm 2.59$ for controls, with mean top-1 energy fractions of
0.695 and 0.340 and mean top-3 fractions of 0.850 and 0.481,
respectively. In other words, coherent NMR variability continues to be
driven by a few dominant distance-space modes, whereas synthetic incoherent
controls distribute variance across many more directions.

The figure also shows that the effective-rank result is not tied to a
single formula. Participation-ratio SCI and entropy-based SCI retain
near-identical orderings, so the discriminative power of the metric is
coming primarily from the distance-variance representation rather than
from one specific spectrum summary.

These results also explain why SCI softens on Main110 without implying
that the spectral signal itself is weak. When we benchmark raw spectral
quantities directly, $r_\text{eff}$, the top-1 energy fraction, and the
top-3 energy fraction all reach grouped and
leave-one-function-class-out AUCs near 0.99, with perfect holdout
discrimination (Table~\ref{tab:rescue}). Likewise, augmenting SCI with
residue count and model count lifts grouped CV AUC from 0.968 to 0.996.
We treat these rescue models as explanatory rather than as replacements
for SCI: they show that the main performance loss on Main110 is driven
by the bounded normalization across heterogeneous $L$ and $M$, not by
disappearance of the underlying spectral separation. We do not recommend
$r_\text{eff}$ or top-energy fractions as primary practical QC scores
because, unlike SCI, they do not provide a shared $[0,1]$-bounded scale
and require recalibration across proteins with different residue counts
and ensemble sizes. SCI therefore trades some discrimination for a
portable, interpretable coherence summary that can be compared across
heterogeneous cohorts on a common scale.

\begin{table}[!t]
\caption{Effective-rank summary by group.}
\label{tab:reff}
\centering
\begin{tabular}{lcccc}
\hline
Group & $r_\text{eff}$ & Top-1 frac. & Top-3 frac. & $n$ \\
\hline
Main110 NMR & $2.13 \pm 0.67$ & 0.695 & 0.850 & 110 \\
Synthetic incoherent controls & $7.93 \pm 2.59$ & 0.340 & 0.481 & 110 \\
AlphaFold proxy & $2.20 \pm 0.44$ & --- & --- & 40 \\
\hline
\end{tabular}
\end{table}

\begin{table*}[!t]
\caption{Spectral rescue analysis. These models are reported to explain
why SCI softens on Main110; they do not replace SCI as the canonical
headline metric.}
\label{tab:rescue}
\centering
\begin{tabular}{lcccc}
\hline
Model & Raw AUC & Grouped CV AUC & LOFO AUC & Holdout AUC \\
\hline
SCI & 0.973 & 0.968 & 0.971 & 0.983 \\
$r_\text{eff}$ & 0.991 & 0.990 & 0.990 & 1.000 \\
Top-1 energy fraction & 0.992 & 0.990 & 0.990 & 1.000 \\
Top-3 energy fraction & 0.992 & 0.991 & 0.991 & 1.000 \\
SCI + $n_\text{models}$ & 0.987 & 0.984 & 0.985 & 1.000 \\
SCI + $n_\text{residues}$ & 0.986 & 0.983 & 0.984 & 0.975 \\
SCI + $n_\text{residues}$ + $n_\text{models}$ & 0.998 & 0.996 & 0.997 & 1.000 \\
\hline
\end{tabular}
\end{table*}

\subsection{Comparison with Baseline Metrics}
\label{sec:res_baselines}

The expanded cohort changes the comparative ranking of metrics. At the
single-metric raw-discrimination level, SCI remained strong
(AUC $= 0.9726$), $\sigma_{R_g}$ was higher
(AUC $= 0.9859$), PCA variance ratio remained competitive
(AUC $= 0.9669$), and contact density stayed near chance
(AUC $= 0.5051$). However, Main110 makes the relative positioning more
explicit than the pilot: across every reported regime,
$\sigma_{R_g}$ is the stronger single-feature discriminator, and the
multifeature QC-full model is strongest overall. We therefore do not
claim that SCI is the empirically best scalar on this benchmark.

\begin{figure}[!t]
\centering
\includegraphics[width=\columnwidth]{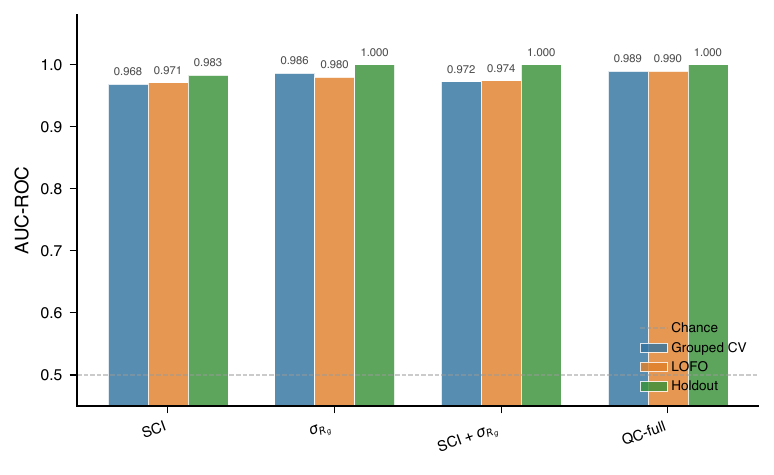}
\caption{Baseline metric comparison. The current manuscript emphasizes
grouped CV, leave-one-function-class-out, and independent holdout
performance for SCI, $\sigma_{R_g}$, SCI+$\sigma_{R_g}$, and QC-full.}
\label{fig:baselines}
\end{figure}

Table~\ref{tab:baselines} therefore summarizes the current predictive
comparison. SCI remained robust under grouped stratified
cross-validation and leave-one-function-class-out testing
(AUC $= 0.9681$ and $0.9707$), confirming that the coherence signal
generalizes beyond individual proteins. Yet the best-performing
predictive model on the expanded cohort was QC-full, which combines SCI,
$\sigma_{R_g}$, PCA variance ratio, and smoothness
(AUC $= 0.9893$ grouped CV; $0.9896$ leave-one-function-class-out).
This shift in ranking is informative. It shows that SCI should
be interpreted as the most transparent coherence axis, not as a claim to
dominate amplitude-sensitive baselines on every discrimination task.
Amplitude and plausibility features carry additional information once
the benchmark is scaled to Main110.
Figure~\ref{fig:baselines} visualizes the same ranking at the level of
the manuscript's three transfer-oriented validation regimes.

\begin{table*}[!t]
\caption{Baseline and multimetric comparison on the expanded cohort.}
\label{tab:baselines}
\centering
\begin{tabular}{lccccc}
\hline
Metric or model & Raw AUC & Grouped CV AUC & LOFO AUC & Holdout AUC & Direction \\
\hline
SCI & 0.9726 & 0.9681 & 0.9707 & 0.9835 & higher $\to$ NMR \\
$\sigma_{R_g}$ & 0.9859 & 0.9860 & 0.9803 & 1.0000 & higher $\to$ NMR \\
PCA $\lambda_1/\Sigma$ & 0.9669 & --- & --- & --- & higher $\to$ control \\
Contact density & 0.5051 & --- & --- & --- & uninformative \\
SCI + $\sigma_{R_g}$ & 0.9837 & 0.9725 & 0.9742 & 1.0000 & multivariate \\
QC-full & 0.9929 & 0.9893 & 0.9896 & 1.0000 & multivariate \\
\hline
\end{tabular}
\end{table*}

\subsection{Robustness Analysis}
\label{sec:res_robust}

\subsubsection{Noise Sweep}

\begin{figure*}[!t]
\centering
\includegraphics[width=\textwidth]{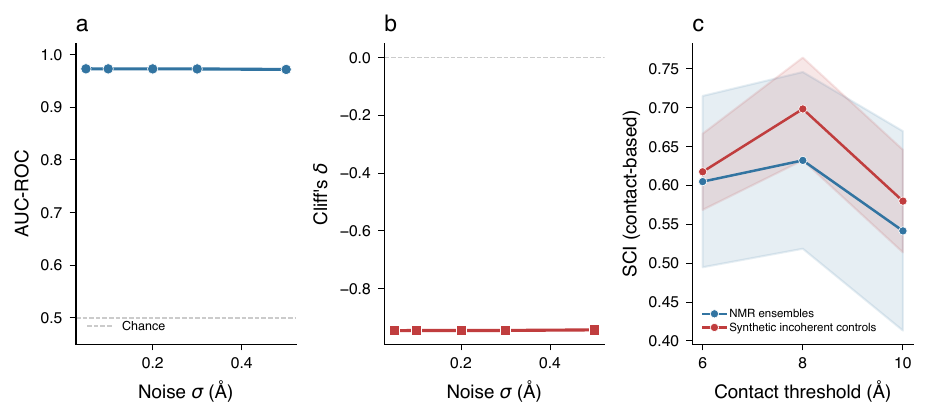}
\caption{Robustness analysis. \textbf{(a)}~AUC-ROC versus injected
noise level. \textbf{(b)}~Cliff's $\delta$ versus injected noise.
\textbf{(c)}~Contact-based SCI versus contact threshold across NMR,
matched synthetic incoherent controls, and AlphaFold references.}
\label{fig:robust}
\end{figure*}

SCI classification performance remained nearly invariant across all
tested noise levels used in synthetic incoherent control generation
($\sigma = 0.05$--$0.50$~\AA): AUC-ROC ranged from 0.9716 to 0.9728,
and Cliff's $\delta$ from $-0.9431$ to $-0.9456$
(Table~\ref{tab:noise}; Fig.~\ref{fig:robust}). The 95\% bootstrap CI
for the mean paired difference narrowed only slightly across the sweep,
consistent with a very modest reduction in the NMR--control gap as the
synthetic incoherent controls gain variance. The headline SCI separation therefore
does not rest on a narrow perturbation setting.

\begin{table}[!t]
\caption{Noise sweep: SCI discrimination across synthetic-incoherent-control noise
levels.}
\label{tab:noise}
\centering
\begin{tabular}{ccccc}
\hline
$\sigma$ (\AA) & AUC & $\delta$ & 95\% CI (paired $\Delta$) & $p$ \\
\hline
0.05 & 0.9728 & $-0.9456$ & $[0.306,\;0.371]$ & $< 10^{-4}$ \\
0.10 & 0.9727 & $-0.9455$ & $[0.305,\;0.371]$ & $< 10^{-4}$ \\
0.20 & 0.9726 & $-0.9453$ & $[0.305,\;0.370]$ & $< 10^{-4}$ \\
0.30 & 0.9726 & $-0.9451$ & $[0.304,\;0.369]$ & $< 10^{-4}$ \\
0.50 & 0.9716 & $-0.9431$ & $[0.301,\;0.367]$ & $< 10^{-4}$ \\
\hline
\end{tabular}
\end{table}

\subsubsection{Youden's $J$ Threshold and Contact-Threshold Sensitivity}

Youden's-$J$ thresholding identifies a practical operating point for the
grouped primary analysis, but the expanded cohort shows that thresholds
should be interpreted as screening heuristics rather than universal
pass/fail rules. The grouped Main110 optimum was $\tau^* = 0.811$ with
95.5\% sensitivity and 89.1\% specificity, while the PDB-level
sensitivity analysis yielded a closely matching optimum
($\tau^* = 0.813$). The contact-based SCI variant retained qualitative
separation between NMR and synthetic incoherent ensembles across multiple contact
cutoffs, but its exact values shifted with threshold choice, reinforcing
why the manuscript anchors its claims on the continuous distance-based
formulation.

\subsubsection{Ensemble-Size Sensitivity}

Subsampling analysis shows that SCI stability depends strongly on the
number of available conformers. The mean absolute deviation from the
full-ensemble SCI estimate shrank from 0.548 at five models to 0.154 at
ten, 0.050 at fifteen, and 0.005 at twenty
(Table~\ref{tab:ensemble_size}). This directly explains why low-model
ensembles are overrepresented among the Main110 NMR threshold failures
and suggests that practical SCI deployment is most stable once roughly
15--20 conformers are available.

\begin{table}[!t]
\caption{Ensemble-size sensitivity of SCI relative to the full-ensemble
estimate.}
\label{tab:ensemble_size}
\centering
\begin{tabular}{cccc}
\hline
Models retained & Mean abs.\ deviation & Median abs.\ deviation & $n$ \\
\hline
5 & 0.548 & 0.515 & 790 \\
10 & 0.154 & 0.135 & 790 \\
15 & 0.050 & 0.043 & 790 \\
20 & 0.005 & 0.000 & 790 \\
\hline
\end{tabular}
\end{table}

\subsection{Exploratory: Biological Validation and Contextual Flexibility}
\label{sec:res_exploratory}

SCI remained biologically grounded after the cohort expansion. Across
110 proteins with residue-level analyses, the median Spearman
correlation between SCI-derived residue contributions and experimental
RMSF was $0.587$, while the median correlation between GNM-predicted
and experimental RMSF was $0.701$. In 108 of the 110 proteins, residue
contributions were positively associated with RMSF. A paired Wilcoxon
comparison rejected equality in the direction GNM $>$ contribution
(one-sided $p = 0.0189$; two-sided $p = 0.0378$), indicating that the
SCI-derived contribution map is biologically meaningful but does not
replace a structure-based elasticity model.

The residue-level outliers are informative
(Table~\ref{tab:residue_outliers}). Only two proteins showed negative
contribution-versus-RMSF associations (1ARK and 1YS5), and three
additional proteins remained weak ($\rho < 0.2$): 1OQY, 2JQF, and 6U1O.
None of these five overlap the grouped SCI threshold-failed set. The
threshold-failed protein 2LM3, by contrast, still showed a positive
residue-level correlation ($\rho = 0.213$), indicating that
ensemble-level screening errors and residue-level biological outliers
are not the same phenomenon. The two negative cases point to different
failure modes. For 1ARK, both SCI contribution and GNM are weak,
suggesting a flexibility pattern that is not well captured by either a
low-rank SCI contribution map or a simple elastic-network description.
For 1YS5, GNM remains moderately aligned with RMSF whereas SCI
contribution is nearly flat, indicating that the residue-level SCI map
and GNM are emphasizing different aspects of the underlying motion.

The pair-based analyses are therefore best interpreted as contextual
case studies rather than as formal validation layers. Across 14
allosteric pairs, the median absolute change in SCI-derived summary was
only 0.00356, whereas the corresponding median absolute changes were
0.208 for contact density and 4.276~\AA\ for radius of gyration. Across
3 apo/holo pairs, the same pattern persisted (median absolute changes
0.000319, 0.0582, and 0.831~\AA, respectively), but the sample is too
small to support formal inference. The 4 available intrinsically
disordered proteins remained highly flexible (median mean RMSF
1.878~\AA) while still showing high single-structure proxy SCI
(median 0.974), reinforcing that coherence and flexibility are not the
same biological dimension.

\begin{table}[!t]
\caption{Lowest residue-level SCI contribution correlations with
experimental RMSF.}
\label{tab:residue_outliers}
\centering
\begin{tabular}{lcccc}
\hline
PDB & $L$ & Contribution vs.\ RMSF & GNM vs.\ RMSF & Note \\
\hline
1ARK & 60 & $-0.505$ & 0.125 & negative \\
1YS5 & 156 & $-0.025$ & 0.493 & negative \\
1OQY & 363 & 0.056 & $-0.632$ & weak \\
2JQF & 166 & 0.081 & 0.239 & weak \\
6U1O & 258 & 0.110 & $-0.329$ & weak \\
\hline
\end{tabular}
\end{table}

\subsection{Eigenvector Smoothness as a Complementary Diagnostic}
\label{sec:res_smooth}

Having established that coherence and amplitude form the two primary QC
axes, we asked whether a complementary spatial-plausibility attribute
adds information not captured by either. The mean top-eigenvector
smoothness score was $0.916 \pm 0.070$ for NMR versus
$0.842 \pm 0.027$ for synthetic incoherent controls, with a moderate-to-large
effect size (Cliff's $\delta = -0.678$) and paired permutation
$p < 10^{-4}$. AlphaFold proxy values were highest
($0.963 \pm 0.035$), consistent with single structures producing very
smooth dominant geometric modes.

Among the Main110 NMR ensembles, smoothness was positively correlated
with SCI (Spearman $\rho = 0.365$, $p < 10^{-4}$) but even more strongly
with $\sigma_{R_g}$ ($\rho = 0.833$, $p \ll 10^{-6}$), suggesting that
it captures partially independent information while remaining closer to
a plausibility check than to a primary discrimination axis. The
SCI\,$\times$\,$\sigma_{R_g}$ map and the smoothness overlays therefore
provide a useful multimetric QC representation, but not a single
universal decision rule.

\begin{figure*}[!t]
\centering
\includegraphics[width=\textwidth]{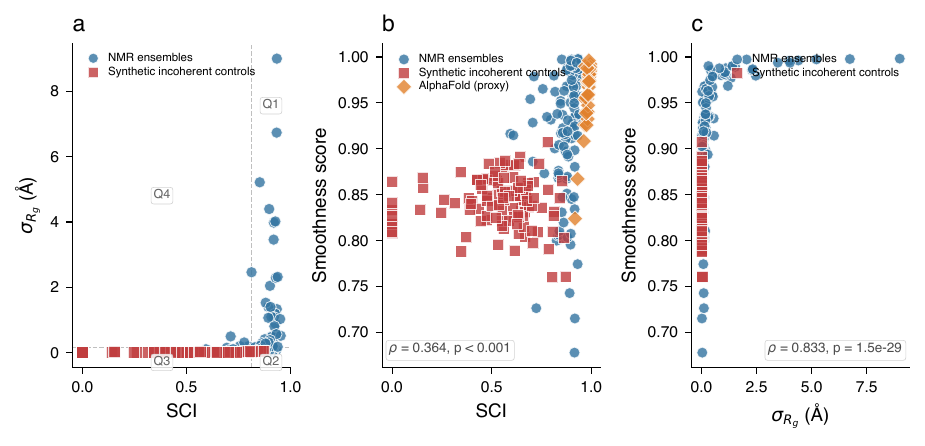}
\caption{Multi-metric QC diagnostics. \textbf{(a)}~SCI versus
$\sigma_{R_g}$ for all samples. \textbf{(b)}~Top-eigenvector smoothness
score versus SCI. \textbf{(c)}~Smoothness versus $\sigma_{R_g}$. The
weak-to-moderate SCI--smoothness correlation indicates partially
independent diagnostic axes.}
\label{fig:qcdiag}
\end{figure*}

\subsection{Failure Mode Analysis}
\label{sec:failure_modes}

\begin{figure*}[!t]
\centering
\includegraphics[width=\textwidth]{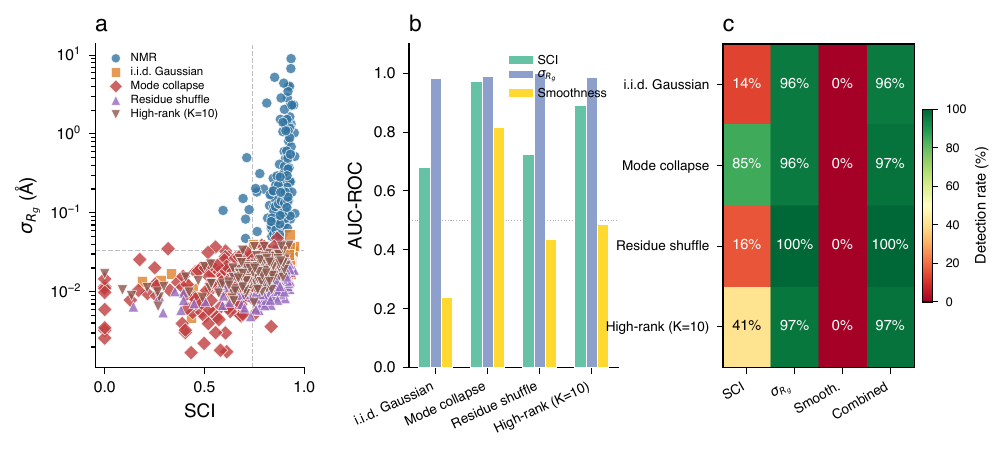}
\caption{Failure mode analysis for the expanded QC stress tests.
\textbf{(a)}~SCI versus $\sigma_{R_g}$ for NMR ensembles and four
synthetic artifact classes, with dashed screening thresholds.
\textbf{(b)}~AUC-ROC per metric per failure type. \textbf{(c)}~Detection
rate heatmap showing individual-metric and combined screening
performance.}
\label{fig:failure}
\end{figure*}

The four synthetic stress tests occupy distinct regions of the
SCI\,$\times$\,$\sigma_{R_g}$ plane and therefore remain useful for
interpreting which metric is carrying each detection decision.
Gaussian and residue-shuffle artifacts tend to preserve deceptively high
SCI while collapsing amplitude, whereas mode-collapse and high-rank
perturbation reduce SCI more strongly. This pattern is why
$\sigma_{R_g}$ and SCI remain complementary even after the expanded
cohort analysis shows that amplitude-aware models can outperform SCI on
some generalization benchmarks. Table~\ref{tab:failure} and
Fig.~\ref{fig:failure} summarize these failure-type-specific detection
patterns.

\begin{table}[!t]
\centering
\caption{Failure mode detection rates. Individual metric detection
rates (percentage of ensembles flagged) and combined QC gate performance
across four synthetic failure types. Expanded-cohort NMR false-positive
rate under the fixed combined thresholds: 10.1\%.}
\label{tab:failure}
\begin{tabular}{lcccc}
\hline
\textbf{Failure type} & \textbf{SCI} & \textbf{$\sigma_{R_g}$} & \textbf{Smooth.} & \textbf{Combined} \\
\hline
i.i.d.\ Gaussian      & 14\% & 96\% & 0\% & 96\% \\
Mode collapse         & 85\% & 96\% & 0\% & 97\% \\
Residue shuffle       & 16\% & 100\% & 0\% & 100\% \\
High-rank ($K\!=\!10$) & 41\% & 97\% & 0\% & 97\% \\
\hline
\end{tabular}
\end{table}

\section{Discussion}
\label{sec:discussion}

\subsection{Methodological Advantages of SCI}

The central contribution of this manuscript is not a new mathematical
definition relative to the internal 27-protein pilot, but a more
demanding validation setting. Main110 shows that the pilot signal
survives broader heterogeneity, while also revealing softer
specificity, function-class variation, and threshold
non-transferability that were invisible at pilot scale. Those findings
are scientifically useful because they clarify where SCI is robust on
its own and where complementary QC axes become necessary.

Within that broader setting, SCI still offers several methodological
advantages over existing ensemble-diversity measures. First, it is
\emph{rotation-invariant}: because SCI operates on pairwise distances
rather than Cartesian coordinates, no structural superposition or
reference frame is required. This eliminates a well-known source of
bias in PCA-based metrics and RMSD
calculations~\cite{diamond1992relationship, kufareva2012rmsd}. Second,
SCI is \emph{model-free}: it requires no force field, energy function,
or normal-mode computation. Third, SCI produces a \emph{single number
on $[0,1]$} with a direct spectral interpretation, providing a bounded,
portable comparison scale across heterogeneous ensembles even though
that normalization can cost some discrimination in broader cohorts.
Fourth, SCI is
\emph{sensitive to local coordinated motions in principle}: unlike
$\sigma_{R_g}$, which captures only global size fluctuations, SCI can
detect coordinated loop rearrangements, domain twists, and hinge
motions that leave overall size weakly changed.

\subsection{Physical Interpretation of SCI Direction}

A key insight from the Main110 results is the relationship between
coordinate-space and distance-space complexity. Low-rank deformations in
coordinate space do not automatically produce low-rank structure in
distance-variance space, because the nonlinear distance function
$D(i,j) = \|\mathbf{r}_i - \mathbf{r}_j\|$ introduces cross-terms that
disperse variance across many eigenvalues. Conversely, coordinated
conformational motions such as hinge bending and domain closure produce
highly correlated distance changes across many residue pairs,
concentrating the distance-variance spectrum into a few dominant modes.
High SCI should therefore be interpreted as \emph{high motion
coherence}, not simply as low dimensionality or low structural
variability. The rescue analyses summarized in Table~\ref{tab:rescue}
sharpen this interpretation. Raw
spectral summaries such as $r_\text{eff}$ and the top-energy fractions
remain near 0.99 AUC on Main110, whereas normalized SCI softens to
0.973. Likewise, adding $n_\text{residues}$ and $n_\text{models}$ to
SCI nearly restores perfect discrimination. The most parsimonious
explanation is that Main110 heterogeneity enters mainly through the
bounded normalization $d = \min(L, M{-}1)$: as protein length and model
count vary more widely, the same underlying spectral concentration is
compressed unevenly. We therefore view the current SCI as an
interpretable normalized score: a bounded, portable coherence summary
whose shared scale facilitates cross-protein comparison, even though
that portability carries a measurable discrimination tradeoff on
heterogeneous cohorts. The rescue models indicate that the underlying
spectral signal remains stronger than the headline SCI alone suggests.

\subsection{Limitations}

Several limitations should be noted. First, although Main110 is
substantially larger than the pilot cohort, it is still modest relative
to the full diversity of NMR structures in the PDB, and the fixed
holdout contains only 11 proteins. Moreover, the holdout spans only
27--75 residues, which samples the small-protein end of the main-cohort
size range and therefore does not directly test external transfer for
large proteins ($L > 100$). Second, the AlphaFold analysis uses
a single-structure proxy that measures fold-geometry concentration
rather than cross-conformer coherence; the weak pLDDT association
reinforces that this result should be interpreted only as a reference
comparison. Third, the synthetic incoherent controls and failure-mode
generators do not span all real-world failure cases. In particular, the
expanded cohort shows that fixed thresholds calibrated on synthetic
artifacts produce a non-negligible NMR false-positive rate (10.1\%), so
they should be treated as screening heuristics rather than universal
accept/reject rules. Fourth, the pair-based allosteric and apo/holo
analyses are descriptive rather than inferential because the available
sample sizes are small. Fifth, this manuscript deliberately leaves
molecular-dynamics benchmarking out of scope; future work should test
SCI on MD trajectories, AlphaFold-derived ensembles, and poorly
converged experimental refinements.

\subsection{Practical Guidance: When SCI Is and Is Not Informative}

SCI measures spectral concentration of the distance-variance matrix. It
is most informative when comparing ensembles generated by the same or
closely related procedures, where a substantially lower SCI flags
incoherent or insufficiently converged variation. The Main110 results
also make clear, however, that SCI should not be treated as a complete
one-number surrogate for ensemble quality.
$\sigma_{R_g}$ is the empirically stronger single-feature discriminator
on this benchmark, and QC-full is strongest overall. Synthetic artifacts
such as i.i.d.\ Gaussian noise can also retain
moderately high SCI despite lacking physically meaningful structural
coordination.

We therefore recommend reporting three complementary quantities:
\begin{enumerate}
\item \textbf{SCI} as the core coherence axis and the most interpretable
      summary of whether variation is organized into a few coordinated
      modes.
\item \textbf{$\sigma_{R_g}$} as an amplitude axis that catches
      low-variation artifacts and, on Main110, provides the strongest
      standalone discrimination.
\item \textbf{Smoothness score} as a spatial plausibility check on the
      dominant mode.
\end{enumerate}
Table~\ref{tab:ensemble_size} shows that the mean absolute SCI deviation
from the full-ensemble estimate shrank from 0.548 at five models to
0.154 at ten, 0.050 at fifteen, and 0.005 at twenty, suggesting that
SCI is most stable once roughly 15--20 conformers are available.
Fig.~\ref{fig:qcbox} summarizes the recommended practical workflow.

\begin{figure}[!t]
\centering
\fbox{\begin{minipage}{0.93\columnwidth}
\small
\textbf{Procedure: Three-Metric Ensemble QC Gate}\\[4pt]
\textbf{Input:} Ensemble $\{S_1, \ldots, S_M\}$ with C$\alpha$
coordinates.\\
\textbf{Output:} QC summary in terms of coherence, amplitude, and
plausibility.\\[4pt]
\textbf{Step 1.} Compute $V(i,j) = \mathrm{Var}_m[D_m(i,j)]$ and derive
$\mathrm{SCI}$ via Eq.~\eqref{eq:sci}.\\
\textbf{Step 2.} Compute $\sigma_{R_g}$ across models.\\
\textbf{Step 3.} Compute top-eigenvector smoothness via the Rayleigh
quotient on $\mathcal{L}$ (Section~\ref{sec:stats}).\\
\textbf{Step 4.} Report the ensemble on the
SCI\,$\times$\,$\sigma_{R_g}$ plane and use smoothness as a supporting
plausibility check.\\[4pt]
\emph{Interpretation:} Fixed thresholds such as
$\tau_\mathrm{SCI} = 0.74$, $\tau_{\sigma_{R_g}} = 0.033$~\AA, and
$\tau_\mathrm{smooth} = 0.75$ are useful screening heuristics for the
synthetic stress tests, but on Main110 they still flag 10.1\% of the
NMR reference set. They should therefore be used for triage rather than
as universal pass/fail rules.
\end{minipage}}
\caption{Recommended QC procedure. SCI and $\sigma_{R_g}$ define the
core 2D detection plane, while smoothness adds a spatial plausibility
check. The fixed thresholds summarize the synthetic stress tests but are
not universal acceptance criteria on Main110.}
\label{fig:qcbox}
\end{figure}

\subsection{Informatics Relevance}

SCI is not intended to explain dynamical mechanisms but to serve as an
automated informatics QC metric for large-scale structural resources and
ensemble-generation pipelines. When ensemble quality is unchecked,
downstream biomedical inferences such as binding-site prediction,
epitope mapping, and variant interpretation may silently propagate
artifacts from incoherent ensembles. The Main110 results suggest that
SCI is most useful when reported with grouped inference, a PDB-level
sensitivity check, and complementary amplitude and smoothness
diagnostics.

\subsection{Future Directions}

Several extensions are natural. SCI could be applied to molecular
dynamics trajectories by treating snapshots as ensemble members,
enabling coherence-aware quality assessment of conformational sampling.
The same framework could be used to compare experimentally derived NMR
ensembles with ensemble-generating predictors such as AF-Cluster or
Boltzmann-style models~\cite{wayment2024predicting, noe2019boltzmann}.
At the residue level, the current biological validation motivates closer
integration of SCI-derived contribution maps with allostery analysis,
disorder characterization, and function-specific classifiers.

\section{Conclusion}
\label{sec:conclusion}

We have proposed the Spectral Coherence Index (SCI), a model-free,
rotation-invariant metric for quantifying how coherently conformational
variation is organized in protein structural ensembles. On the expanded
Main110 cohort, SCI robustly separates experimental NMR ensembles from
matched synthetic incoherent controls at the grouped primary level
(AUC $= 0.9726$, Cliff's $\delta = -0.945$), remains stable at the
PDB level (AUC $= 0.9716$), and transfers to an independent 11-protein
holdout (AUC $= 0.9835$).

The expanded benchmark clarifies the role of SCI in a modern
ensemble-QC stack. $\sigma_{R_g}$ is the stronger empirical
single-feature discriminator on Main110, while SCI remains the most
interpretable single coherence signal, with strong grouped and
leave-one-function-class-out generalization. Rescue analyses show that
the underlying spectral separation remains very strong and that much of
the observed softening comes from normalization across heterogeneous
residue counts and model counts rather than from loss of the spectral
signal itself. A QC-augmented model achieves the best predictive
performance by combining SCI with amplitude and plausibility features.
Residue-level biological validation links SCI-derived contributions to
experimental RMSF and GNM predictions, and the AlphaFold reference
analysis supports the spectral formalism while emphasizing that
single-structure proxies are not substitutes for ensemble coherence.

Overall, SCI should be viewed as a principled coherence axis for
protein-ensemble informatics: not the empirically strongest standalone
discriminator on Main110, but the most transparent bounded summary of
coherent distance-space organization. Its practical value lies in
providing a portable comparison scale across heterogeneous ensembles,
even when that portability entails a measurable discrimination tradeoff,
and it is therefore most useful when embedded in grouped evaluation and
a multi-metric QC workflow.

\end{document}